\documentclass[twocolumn,showpacs,preprintnumbers,amsmath,amssymb]{revtex4-1}
\usepackage[dvipdfmx]{graphicx}
\usepackage{dcolumn}
\usepackage{bm}
\usepackage{color}
\usepackage{here}
\usepackage{comment}
\usepackage{mathcomp}

\newcommand{\be}{\begin{eqnarray}}
\newcommand{\ee}{\end{eqnarray}}

\newcommand{\bc}{}

\begin{document}

\title{Critical velocity for wake vortex generation behind a plate in a superflow}
\author{Haruya Kokubo$^{1}$, Hiromitsu Takeuchi$^{2,3}$ and Kenichi Kasamatsu$^{1}$}
\affiliation{${}^1$Department of Physics, Kindai University, Higashi-Osaka, Osaka 577-8502, Japan\\ ${}^2$ Department of Physics, Osaka Metropolitan University, 3-3-138 Sugimoto, Osaka 558-8585, Japan \\
${}^3$ Department of Physics, Nambu Yoichiro Institute of Theoretical and Experimental Physics (NITEP), Osaka Metropolitan University, 3-3-138 Sugimoto, Osaka 558-8585, Japan
}

\date{\today}

\begin{abstract}
We study theoretically the critical velocity $U_c$ for quantum vortex generation by a thin plate-shaped obstacle moving through a uniform Bose-Einstein condensate. 
Our results based on the Gross-Pitaevskii theory reveal that the critical velocity monotonically decreases with increasing plate size $L$.
In the limit of large $L$, the critical velocity is asymptotic to $L^{-1/2}$ predicted by the potential flow theory for an incompressible ideal fluid with a phenomenological length correction. 
As $L$ decreases, however, the incompressible analysis breaks down quantitatively.
By performing a perturbative analysis to incorporate compressibility into the potential flow theory, we have successfully reproduced the numerical results analytically over a wide parameter range.
It is also shown that the critical velocity increases with finite plate thickness.
\end{abstract}

\pacs{
} 
\maketitle

\section{Introduction}
When an obstacle moves within a superfluid at speeds exceeding the critical velocity, dissipation can arise as a result of generation of elementary excitations and quantum vortices. 
Vortex nucleation and subsequent wake dynamics formed by a moving obstacle in superfluids has been experimentally studied in cold atomic-gas Bose-Einstein condensates (BECs) \cite{Raman1999-rb,Onofrio2000-wc,Kwon2015-do,Kwon2015-kw,Kwon2016-cc,Lim2022-uw} and atomic-gas Fermi superfluids \cite{Weimer2015-lu,Park2018-ne,Kwon2021-oz}.
The critical velocity is the most fundamental issue among the related topics, having been extensively studied both 
experimentally \cite{Raman1999-rb,Kwon2015-do,Weimer2015-lu,Park2018-ne} and theoretically \cite{Feynman1955-cj,Frisch1992-ul,Jackson1998-ct,Nore2000-zt,Winiecki2000-wc,Rica2001-tr,Rica2001-cb,Aftalion2003-xn,Varoquaux2006-bb,Piazza2013-kf,Pinsker2014-el,Stagg2014-to,Kunimi2015-nk,Kiehn2022-xk,Muller2022-mf,Kwak2023-la,Kokubo2024-jf,Huynh2024-wv}. 
According to the Landau criterion for superfluidity in weakly interacting Bose gases, the critical velocity in a uniform system is known to match the sound velocity $c_s$. 
However, when an obstacle potential is present within the BEC, the critical velocity is suggested to be lower than the sound velocity and to depend on the details of the obstacle in non-trivial manners \cite{Feynman1955-cj,Raman1999-rb,Rica2001-cb,Rica2001-tr,Kwon2015-do,Kwak2023-la}.

The main causes of the decrease in critical velocity is (i) the local velocity near the obstacle can be larger than the global velocity, and (ii) the sound velocity can be spatially variable due to the density inhomogeneity in the local density approximation. 
Within the Gross-Pitaevskii (GP) model, the critical velocity for a cylinder potential is $0.37 c_s$ when its radius is much larger than the healing length $\xi$ \cite{Rica2001-tr,Pham2005-uh}.
These theoretical predictions have been tested by the experiment \cite{Kwon2015-do}, which has demonstrated the nontrivial size dependence of the obstacle potential for the critical velocity. 
In cold atomic BECs, the obstacle made of the laser beam, whose intensity decays away from the center, has a profile modeled generally by a Gaussian function.
In this case, the shape of the obstacle is determined by two parameters—the width and height of the Gaussian function, which define the effective size of the obstacle. 
However, even with the same effective obstacle size, the profile of the region where the potential height is lower than the chemical potential varies depending on the parameters. 
Since this Gaussian tail affects the critical velocity, it becomes being difficult to accurately understand the size dependence of the obstacle.
This represents a crucial issue that must be addressed for achieving a universal classification of superfluid wake flows \cite{Kwon2015-do,Kwak2023-la,Kokubo2024-jf}.
Recently, Huynh \textit{et al.} have reported the analytical calculation of the critical velocity for the Gaussian obstacles, but the size dependence of the potential has not been discussed explicitly \cite{Huynh2024-wv}.

To overcome the difficulty inherent in describing the scale dependence of the superfluid critical velocity, we discuss a fundamental case of the superfluid wake behind a hard-core `plate-shaped' obstacle potential placed in the two-dimensional (2D) uniform BEC.
Recent developments in digital micro-mirror devices in experiments on cold atomic systems have made it possible to form plate-shaped barriers \cite{Gauthier2019-nq}.
When the thickness of the plate is much smaller than the healing length, the size of the plate-shaped potential can be characterized only by the length of the plate.
The single-parameter characterization enables us to make a rigid analysis of the size dependence of the critical velocity.
We first calculate the critical velocity by use of simulations of the GP equation and show that it monotonically decreases with increasing the plate size.
To support the numerical results we analytically estimate the critical velocity using potential flow theory for fluid flow in a 2D space, where a length correction is introduced to incorporate the quantum pressure phenomenologically. 
Here, we employ the Joukowski transformation from the well-known flow around a cylinder to derive the flow profile around a plate.
By including compressibility in a perturbative manner, we succeed to explain the size dependence of the critical velocity analytically down to the size as small as the healing length.
Although there has been no clear answer to the size dependence of the critical velocity, except for a simple case such as a disk-shaped obstacle, our study provides one of the few cases where the size dependence can be successfully predicted analytically.

The paper is organized as follows.
In Sec.\ref{sec:1}, we describe the stationary flow around plate-shaped obstacle in a 2D uniform BEC. 
In Sec.\ref{sec:2}, we discuss the size dependence of the critical velocity for a plate-shaped obstacle through the numerical calculations of the GP equation. 
The numerical results are explained on the basis of the analysis in Sec. \ref{sec:1}.
The critical velocity for a plate with finite thickness is also discussed here.
Section \ref{sec:conclusion} devotes to the conclusions and discussion.

\section{Stationary flow}\label{sec:1}
In this section, we first introduce the formulation to consider the problem of the wake in a uniform superfluid system using the 2D GP equation. 
In the incompressible limit and if the flow is laminar, the profile of the superflow velocity field is described analytically by the complex potential flow theory. 
The flow around the plate-shaped obstacle can be derived by the coordinate transformation, known as the Joukowski transformation, from the flow around the cylinderical object. 
The velocity distribution predicted here will be used for estimating the critical velocity of the superfluid in Sec. \ref{sec:2}.  

\subsection{Basic formulation}
We consider that a BEC consisting of atoms with a mass $m$ is infintely extended in the transverse ($xy$-)plane and is strongly confined in the longitudinal direction.
We introduce a static obstacle potential $V(x,y)$ in a uniform BEC.
The dynamics of the wave function $\Psi(\bm{r},t)$ can be described by the time-dependent GP equation 
\begin{equation}
    i\hbar\frac{\partial}{\partial t}\Psi=\left[-\frac{\hbar^2}{2m}\nabla^2+V(x,y)-\mu+g|\Psi|^2\right]\Psi.
    \label{eq:gpe}
\end{equation}
Here, $\mu$ is the chemical potential and $g$ is the coupling constant in the 2D system.
In the following presentation of the results, length, time, and energy are scaled with $\xi=\hbar/\sqrt{2m\mu}$, $\omega^{-1}=\hbar/\mu$ and $\mu$, respectively.
The sound velocity is written by $c_s=\sqrt{\mu/m} = \sqrt{2} \xi \omega$. 
By substituting the wave function $\Psi = \sqrt{n} e^{i \theta}$, expressed in terms of the condensate density $n$ and the phase $\theta$, into Eq.\eqref{eq:gpe}, we obtain
\begin{equation}
    \frac{\partial}{\partial t}n+\nabla\cdot(n\bm{v})=0,
    \label{eq:fld}
\end{equation}
\begin{equation}
    m\frac{\partial}{\partial t}\bm{v}=-\nabla\left(gn+V-\frac{\hbar^2}{2m\sqrt{n}}\nabla^2\sqrt{n}+\frac{mv^2}{2}\right) ,
    \label{eq:euler}
\end{equation}
where the superfluid velocity is given as the form of the  potential flow  $\bm{v}=(\hbar/m) \nabla\theta$.
The obstacle potential $V$ is assumed to be a hard-core potential and will be incorporated as a boundary condition; it is thus not indicated explicitly in the formulation of this section.
Equation \eqref{eq:fld} is the usual continuity equation, while Eq.\eqref{eq:euler} corresponds to the Euler equation for inviscid compressible flow, where the additional quantum pressure term $\hbar^2 \nabla^2 \sqrt{n}/ (2 m \sqrt{n})$ is included. 
When the spatial scale of flow variation is much larger the intrinsic microscopic scale, namely the healing length $\xi$, the quantum pressure term can be neglected. 
In the following analysis using the hydrodynamic formulation, it is reasonable to neglect quantum pressure term since we confine ourselves to the vortex-free flow.

By neglecting the quantum pressure term the considering a steady state, Eq.\eqref{eq:euler} reduces to the Bernoulli's relation 
\begin{equation}
    gn(\bm{r}) + \frac{1}{2}mv(\bm{r})^2 = gn_0 + \frac{1}{2}mU^2 = \mathrm{const.}\ ,
    \label{eq:dens_Bern}
\end{equation}
where we assume that the density and velocity of the uniform BEC far from the obstacle potential are given by $n_0 \equiv \mu/g$ and $U$, respectively.
Substituting Eq.\eqref{eq:dens_Bern} into Eq.\eqref{eq:fld} for the steady state, we obtain
\begin{equation}
    \nabla\cdot\left\{\left[\frac{1}{M^2} + \frac{1}{2}\left(1 - \frac{v^2}{U^2}\right)\right]\bm{v}\right\} = 0.
    \label{eq:rica-a}
\end{equation}
Here, $M$ is the Mach number, the ratio of the bulk velocity to the sound velocity:
\begin{equation}
M = \frac{U}{c_s}.
\end{equation}
Equation \eqref{eq:rica-a} will be used to evaluate the impact of compressibility in the stationary flow.

\subsection{Potential flow theory}\label{sec:1b}
In classical hydrodynamics, the complex potential theory is usually used to analyze the 2D vortex-free flow of an incompressible ideal fluid. 
When the velocity of a uniform superflow is slower than the critical velocity, the flow can be described by the vortex-free flow of an ideal fluid.
Thus, we use the complex potential theory for description of the stationary flow around the obstacle potential in the superfluid.
The complex potential theory has been used in the analysis of the critical velocities for elliptical and disk obstacles \cite{Stagg2014-to,Rica2001-cb}.
We also incorporate compressibility effects through a perturbative approach, known as $M^2$-expansion \cite{Imai1957-ee}. In this method, the complex velocity potential is still well-defined, so that the methods of complex function theory are applicable. We adapt it to analyse the flow around the plate-shaped obstacle by including the compressible contribution.

Let us first consider the incompressible limit with $n=\mathrm{const.}$, which is valid when the flow velocity is sufficiently slower than the sound velocity. 
Equation \eqref{eq:rica-a} in the limit of $M \to  0$ gives $\nabla \cdot \bm{v} = 0$, which then implies the presence of the stream function $\sigma$ satisfying $v_x = \partial_y \sigma$ and $v_y = -\partial_x \sigma$. 
Also, the velocity potential $(\hbar/m)\theta \equiv \Theta$ obeys the Laplace equation $\nabla^2 \Theta = 0$.
In this case, the Cauchy-Riemann equation is established between $\sigma$ and $\Theta$, and one can set a regular function $W=\Theta+i\sigma$ in the complex space $z=x+iy$.
Thereby, a theoretical treatment in complex function theory allows for the analysis of vortex-free flow in arbitrary incompressible perfect fluids.
By using transformations on the complex plane, e.g. the Joukowski transformation, the flow around obstacles of various shapes can be described.
The Joukowski transformation is a mapping between the $z$- and $z'$-plane through the relation $z' = z + R^2/z$.
By using this transformation, the region outside a circle with radius $R$ centered at the origin in the $z$-plane is mapped to the entire $z'$-plane excluding a line segment of length $4R$ on the real axis.
Here, we use the Joukowski transformation \cite{acheson1990-efd} for the flow around a cylinder to derive the flow around a plate-shaped obstacle.

In the complex plane $z = x + iy$, the velocity potential representing a flow in the $y$-direction around a cylindrical disk with radius $R$ can be expressed using the complex functions as 
\begin{equation}
    \Theta_0^{\mathrm{d}} = -\frac{iU}{2}\left(z - \frac{R^2}{z}\right)+\mathrm{c.c.}.
    \label{eq:dp_qc_0}
\end{equation}
Differentiating the velocity potential $\Theta$ with complex coordinates $z$ and $z^\ast$ yields the velocity $\bm{v}=(v_x,v_y)=\left( (\partial_z + \partial_{z^\ast} ) \Theta, i (\partial_z  - \partial_{z^\ast} ) \Theta \right)$.
Here, $v_x$ and $v_y$ denote the velocities in the $x$- and $y$-directions, respectively.
The velocity $\bm{v}^{\mathrm{d}}=(v_x^\text{d}, v_y^\text{d})$ around the circle is written by
\begin{equation}
    \bm{v}^{\mathrm{d}}=\left(\frac{iU(z^{2}-z^{*2})}{2z^2z^{*2}},U+\frac{UR^2(z^2+z^{*2})}{2z^2 z^{*2}}\right) .
\label{eq:vel_disk}
\end{equation}
At a point far enough from the cylindrical disk, the flow is uniform in the $y$-direction  with velocity $\bm{v}^{\mathrm{d}}=(0,U)$.
The velocity at the lateral point $z=\pm R$ of the cylinder is $\bm{v}^{\mathrm{d}}=(0,2U)$.
For a circle of radius $R$ in the complex plane, applying the Joukowski transformation $z' = z+ R^2/z$ results in a flat plate of length $4R$. 
In order to consider a plate with length $L$, we set $R = L/4$.
The velocity potential of the flow around the plate-shaped obstacle of the length $L$, transformed by the Joukowski transformation to Eq.\eqref{eq:dp_qc_0}, is described by 
\begin{equation}
    \Theta_0^{\text{p}} = -\frac{iU}{2}\left(\sqrt{z'^2-(L/2)^2}-\mathrm{c.c.} \right).
\label{eq:cvp_panel}
\end{equation}
The velocity field around the plate $\bm{v}^\text{p} = (v^\text{p}_x, v^\text{p}_y)$ is 
\begin{align}
v^\text{p}_x &= -\frac{iU}{2}\left(\frac{z'}{\sqrt{z'^2-(L/2)^2}}-\mathrm{c.c.}\right), \nonumber \\
v^\text{p}_y &= \frac{U}{2}\left(\frac{z'}{\sqrt{z'^2-(L/2)^2}}+\mathrm{c.c.}\right).
\label{eq:cv_panel}
\end{align}
If $v/U$ is large enough compare to unity, we need to consider, namely the spatial variation of the density. 
For stationary flow, Eq.\eqref{eq:rica-a} can be rewritten with the complex coordinates as \cite{Rica2001-tr}
\begin{equation}
\frac{\partial^2 \Theta}{\partial z \partial z^{\ast}} 
= \frac{M^2}{4} \frac{\partial}{\partial z} \left[ \left( \frac{4}{U^2} \left| \frac{\partial \Theta}{\partial z}\right|^2 -1 \right) \frac{\partial \Theta}{\partial z^\ast} \right] + \text{c.c.}.   \label{bibunkeigg}
\end{equation} 
The boundary condition of the cylindrical disk is given by 
\begin{equation}
z \frac{\partial \Theta}{\partial z} + z^\ast \frac{\partial \Theta}{\partial z^\ast}=0 \quad \text{at} \quad |z|=R.
\label{eq:bound}
\end{equation}
To evaluate the compressibility in a perturbative way, we express Eq.~\eqref{bibunkeigg} by the integral form \begin{equation}
\Theta(z, z^*) = \Theta_h(z) + \frac{M^2}{4} \int dz^* \left( \frac{4}{U^2} \left| \frac{\partial \Theta}{\partial z} \right|^2 - 1 \right) \frac{\partial \Theta}{\partial z^\ast} +\mathrm{c.c.}.
\label{eq:rica_pot}
\end{equation}
Here, $\Theta_h(z)=\sum_{n=1}R^{n+1}/z^n$ is determined by the boundary condition of Eq.\eqref{eq:bound} . 
Now, we expand the velocity potential as
\begin{equation}
\Theta = \Theta_0^\mathrm{d} + M^2 \Theta_1^\mathrm{d} + M^4\Theta_2^\mathrm{d} + \cdots.
\label{eq:pot_tenkai}
\end{equation}
The zeroth-order term corresponds to Eq.\eqref{eq:dp_qc_0} in the incompressible limit. 
Substituting Eq.\eqref{eq:dp_qc_0} into Eq.\eqref{eq:rica_pot} we can find the first order correction as
\begin{equation}
\Theta_1^{\mathrm{d}} = iU\frac{ R^6-R^4z^*(6z+z^*)+R^2z z^{*2} (13z-3z^*) }{24z^3 z^{*2}} + \mathrm{c.c.} .
    \label{eq:dp_qc_1}
\end{equation}
The potential $\Theta_0^\text{d} + M^2 \Theta_1^\text{d}$ can be transformed using the Joukowski transformation with $R = L/4$ to yield the potential including compressibility around a plate of length $L$ \cite{Imai1940-cf}.
The corresponding $\Theta_1^{\mathrm{p}}$ is given by
\begin{align}
    \Theta_1^\mathrm{P} = & \frac{iU}{8f} \left( \frac{L}{2} \right)^2 \left[ \frac{13}{6}-\frac{(L/2)^2}{ff^*}+\frac{(L/2)^4}{6f^2f^{*2}} \right] \nonumber \\
&    -\frac{iU}{48f^3} \left( \frac{L}{2} \right)^2 \left[ -\frac{ff^*}{16}+\left( \frac{L}{2} \right)^2 \right] -\mathrm{c.c.} ,
    \label{eq:pp_qc_1}
\end{align}
where we define $f(z') \equiv z' + \sqrt{z'^2 -(L/2)^2 }$.
By repeating the similar calculations, we can evaluate the higher-order correction due to the compressibility. 

\section{Critical velocity}\label{sec:2}
In this section, we discuss the critical velocity $U_c$ associated with the vortex nucleation by the plate-shaped obstacle and its dependence on the plate size $L$.
We first give results of the critical velocity through the numerical calculation of the GP equation.
Then, we compare the numerical results with our analytical estimation based on the potential flow theory.
Finally, we discuss the effect of the plate thickness on the critical velocity. 

\subsection{Numerical analysis}\label{sec:2a}
First, we calculate the critical velocity through numerical calculations.
In the previous section, we discussed a setting in which a fluid flows around a static object. 
In this section, according to the Galilean invariance, we may consider an alternative situation in which an object moves through a static fluid as an equivalent setup. 
Here, we consider the GP equations in a reference frame comoving with the potential $V_p$ as
\begin{equation}
    i\hbar\frac{\partial}{\partial t}\Psi=\left[-\frac{\hbar^2}{2m}\nabla^2-i \hbar U \frac{\partial}{\partial y}+V_{\rm p}(x,y)-\mu+g|\Psi|^2\right]\Psi.
    \label{eq:fgpe}
\end{equation}
The plate-shaped obstacle potential $V_{\rm p}$ is given by 
\begin{equation}
    V_\textrm{p}(x,y)=\begin{cases}V_0 & y=0, \quad |x|\leq L/2\\0 & \textrm{otherwise}\end{cases}.
\end{equation}
In this study, the critical velocity of vortex formation is obtained numerically by using the imaginary time propagation method for the GP equation \eqref{eq:fgpe}, where $t$ is replaced by $-i\tau$.
\begin{figure*}[ht]
    \centering
    \includegraphics[width=\linewidth]{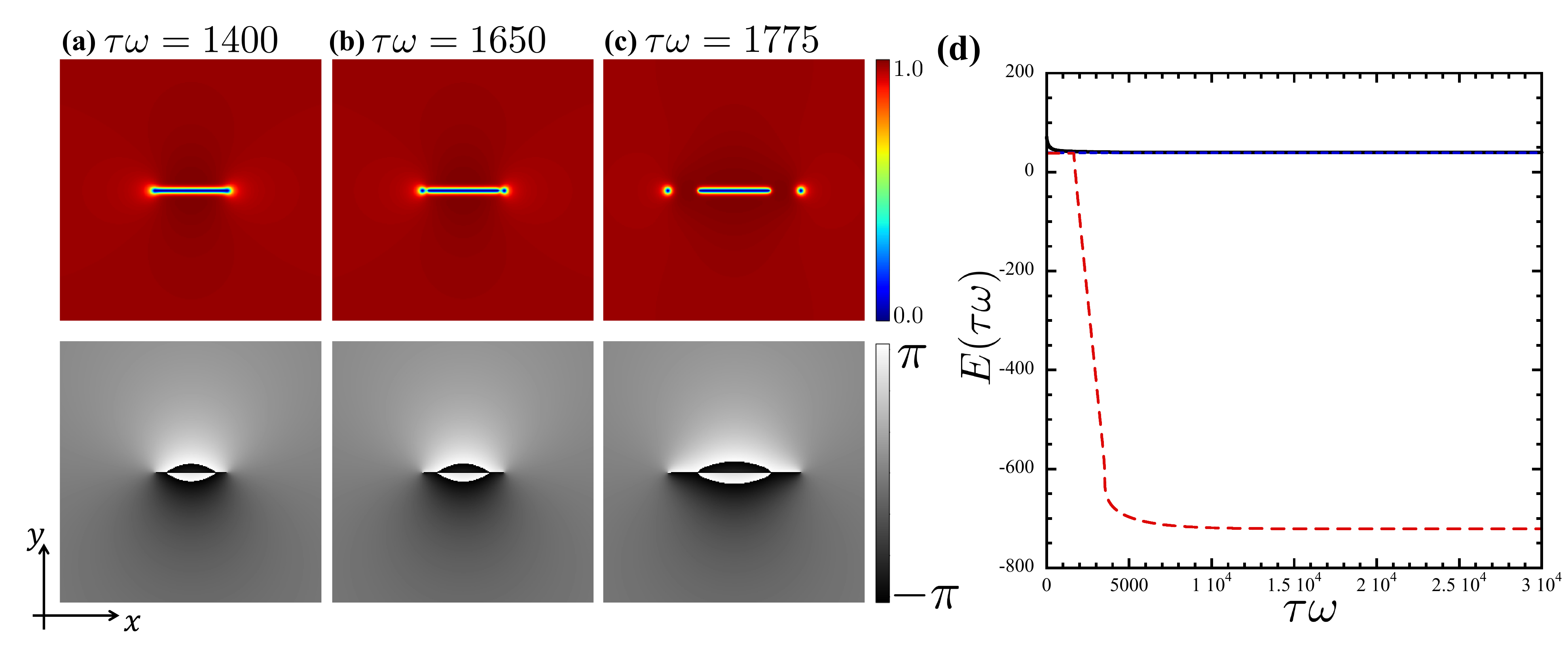}
    \caption{[Panel(a)-(c)] Snapshots of the density profile (top) and the phase profile (bottom) of the condensate wave function during the imaginary time development for $L=40\xi$ and $v=0.226c_s$.
    The spatial region of the plot is $-75\xi<x<75\xi$ and $-75\xi<y<75\xi$.
    The vortices are emitted from the edge of the plate in the extended direction of the plate, and these vortices disappear at the boundary.
    [Panel(d)] Imaginary time evolution of the total energy for an obstacle size of $L=40$. The solid line, dotted line, and dashed line represent the results for $U/c_s=0.223, 0.225,$ and $0.227$, respectively. 
    For $U/c_s=0.223$ and $0.225$, it is observed that the energy decreases due to acceleration and then quickly converges. 
    In contrast, for $U/c_s=0.227$, which exceeds the critical velocity, the energy rapidly decreases due to vortex shedding. From these results, the critical velocity is determined to be $U_c/c_s=0.226$.}
    \label{fig:dens_phase}
\end{figure*}
During imaginary time propagation, the energy of the system monotonically decreases and quantum vortices are formed by the energetic instability.
Since this method is associated with the steepest gradient of the GP energy functional 
\begin{multline}
    E(\tau)=\int d^2 r \left[-\frac{\hbar^2}{2m}\Psi^*\nabla^2\Psi-i \hbar U \Psi^*\frac{\partial}{\partial y}\Psi\right.\\
    \left.+V_\text{p} |\Psi|^2+\frac{g}{2}\left(|\Psi|^2-\frac{\mu}{g}
\right)^2\right],
\end{multline}
it determines the critical parameter value at which the energy barrier sustaining the metastable vortex-free states disappears.
Thus, this method provides a theoretical upper bound of the critical velocity.
In real-time dynamics, the energy difference between the two states is consumed by exciting collective fluctuations, i.e., emitting sound waves.
If such fluctuations exist, vortex nucleation can occur below the critical velocity obtained by imaginary time propagation.

In our simulations, the system size is $400\xi \times 400\xi$, the periodic boundary conditions are imposed in both the $x$- and $y$-directions and the height of the potential of the plate obstacle potential is $V_0 = 100\mu$.

For a hard-core obstacle with $V_0 \gg \mu$, quantum vortices form at the boundary with the obstacle \cite{Kwak2023-la}.
At first of the imaginary time propagation, branch cuts in the phase profile appear in the region just upon the plate-shaped obstacle, where the condensate density is almost absent.
Such branch cuts reach the edge of the plate, developing into the vortex cores [Fig.\ref{fig:dens_phase}(a)] and being subsequently emitted into the condensate as quantum vortices [Fig.\ref{fig:dens_phase}(b)].
Then, the quantum vortices move in a direction parallel to the plate after emission to undergo pair annihilation at the periodic boundary [Fig.\ref{fig:dens_phase}(c)], which corresponds to the phase slip phenomenon leading to the decay of superflow.

We determine the critical velocity by the following steps.
First, we calculate the stationary state at velocity $U=U_0$, which is slightly lower than the expected value. 
When the obstacle velocity is smaller than the critical velocity, the energy decreases slightly and keeps nearly the initial value shown in Fig.\ref{fig:dens_phase}(d).
Next, we increase the velocity by a small amount $\Delta U$ from the steady-state solution at $U=U_0$ and check whether the total energy converges as or not.
When the velocity of the obstacle exceeds the critical velocity, the energy abruptly decreases due to vortex shedding, as indicated by the dashed line in Fig.\ref{fig:dens_phase} (d).
If such a decrease does not occur during a sufficiently long propagation, the velocity is further increased by $\Delta U$ to check the stability of the solution.
Using the velocity $U_{bc}$, at which the energy decreases abruptly, the critical velocity $U_c$ is defined as $U_c = U_{bc} - 0.5\Delta U$. 
Here, the acceleration of the velocity is $\Delta U = 0.002(\xi\omega)$.

\begin{figure}[ht]
    \centering
    \includegraphics[width=\linewidth]{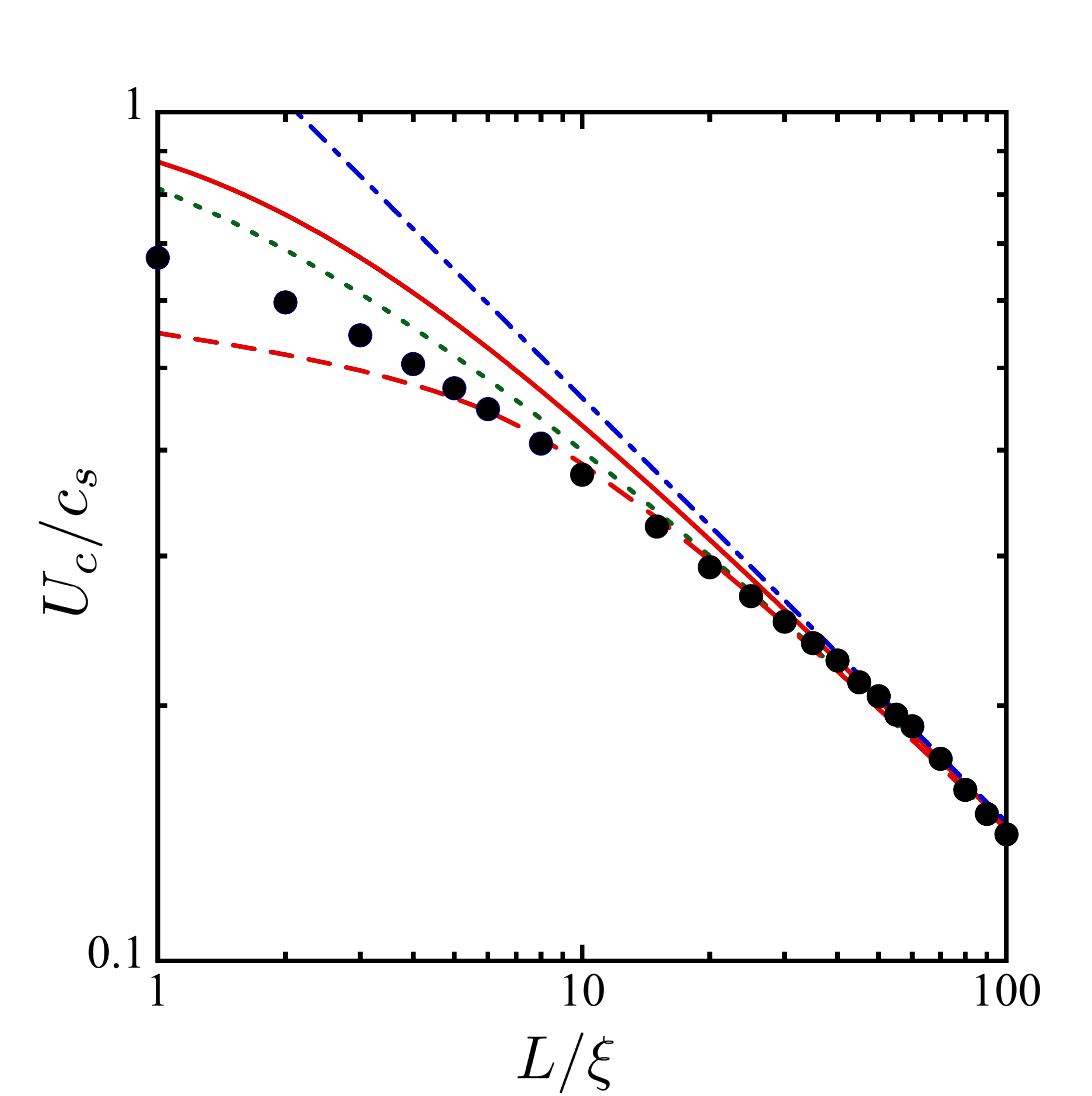}
    \caption{
    The size dependence of the critical velocity of a wake superflow behind the plate-shaped obstacle. 
    The plot show the numerical results for the mesh size $\Delta x,\Delta y=0.5\xi$. 
    The red solid line represents the critical velocity derived from the potential flow theory in the incompresible limit, given by Eq.\eqref{eq:vc_p1} with $l=l_\xi$. When the obstacle size $L$ is sufficiently large, $U_c/c_s$ follows $L^{-1/2}$ (blue dashed-dotted line). 
    The (green) dotted and (red) dashed curves represent the critical velocity by taking into account the compressibility in a perturbative way up to $\mathcal{O}(M^2)$ and $\mathcal{O}(M^{30})$ in the velocity potential, respectively }
    \label{fig:result_1}
\end{figure}

The plots in Fig.\ref{fig:result_1} show the results of the numerical calculations.
The critical velocity decreases monotonically with increasing the plate size $L$, being asymptotic to the power law $L^{-1/2}$ for large $L$.
Moreover, in the region where $L$ is small, the critical velocity deviates from the $L^{-1/2}$ scaling and follows a power law with a lower exponent.
We note that the limit of a point-like object as $L \to 0$, the critical velocity approaches the sound velocity, as has been shown in the previous work \cite{Pham2005-uh}.
We confirm that, if the mesh size $\Delta x, \Delta y$ is less than the healing length $\xi$, obtained results show no quantitative differences, except for the case of the very small size $L \sim \xi$; for example, $U_c/c_s = 0.674$ for $\Delta x(y) = 0.5\xi$ and $U_c/c_s = 0.704$ for $\Delta x (y) = 0.25\xi$.

\subsection{Theoretical analysis}\label{sec:2b}
Next, we analytically examine the size dependence of the critical velocity using the complex potential theory.
A quantitative description of the critical velocity requires an analysis by phenomenologically incorporates the quantum pressure, although it was neglected to analytically describe the stationary flow in Sec.\ref{sec:1}.
To this end, we assume that the formation of quantum vortices in the wake flow behind a plate-shaped obstacle occurs when the local velocity $\bm{v}^\text{p} (x,y)$ reaches the sound velocity at a point $l$ away from the tip of the plate in the extended direction of the plate.
The reason for putting this assumption is as follows.
Near the tip of the plate where vortices are generated, the local velocity increases (actually diverges at a point of the edge), which cause the strong suppression of the local density around the edge. 
Then, our complex potential theory assuming weak compressibility is not applicable. 
When the velocity at a distance comparable to the healing length $(l \sim \xi)$ from the plate reaches the sound velocity, the energy barrier against vortex emission could disappear.
Below we consider $l$ as a fitting parameter to describe the numerical result.

First, we consider the incompressible limit. 
The local velocity at the position of the vortex nucleation can be obtained by substituting $z'=L/2+l$ into Eq.~\eqref{eq:cv_panel}, yielding 
\begin{equation}
v^\text{p}_y = U \frac{L/2+l}{\sqrt{Ll+l^2}}.
\end{equation}
Because of $v_x^\text{p}=0$, we take only the $y$-component of the velocity.
When this local velocity reaches the sound velocity $c_s$, the corresponding bulk velocity $U$ is considered as the critical velocity, being given by
\begin{equation}
   \frac{U_c}{c_s} =\frac{\sqrt{Ll+l^2}}{L/2+l}.
    \label{eq:vc_p1}
\end{equation}
When $L$ is sufficiently large, the critical velocity in Eq. \eqref{eq:vc_p1} can be apporoximated as
\begin{equation}
    \frac{U_c}{c_s}=2\left(\frac{L}{l}\right)^{-1/2}\quad(\mathrm{for}\  L/l \to \infty).
    \label{eq:vc_p1_asy}
\end{equation}
This asymptotic form shows a relationship equivalent to the numerical results, where the critical velocity $U_c$ is proportional to $L^{-1/2}$. 
We treat the distance $l$ as a fitting parameter to compare this asymptotic form with the numerical results for $L > 40$.
The fitting parameter $l$ is determined as
\begin{equation}
    l=0.531 \xi \equiv l_{\xi},
\end{equation}
which is on the order of the healing length as expected.
The red solid curve in Fig.\ref{fig:result_1} represents the result of substituting the estimated fitting parameters into Eq.\eqref{eq:vc_p1}. While the curve describes well the numerical results in the range $L > 20$, there is a discrepancy for smaller obstacles in the range of $L < 20$.

This discrepancy could be due to neglecting the compressibility, which arises as the Mach number $M$ increases. 
We next consider the effect of compressibility by following the perturbative formulation in Sec.\ref{sec:2b}. 
For example, within the leading order correction of $\mathcal{O}(M^2)$ the velocity potential is given by the sum of Eqs.\eqref{eq:cvp_panel} and \eqref{eq:pp_qc_1} as $\Theta^\mathrm{P} = \Theta^\mathrm{P}_0 + M^2\Theta^\mathrm{P}_1$. 
Then, we determine the local velocity at a point separated by a distance $l$ from the edge of the plate, similar to the analysis in Eq.\eqref{eq:vc_p1}, and define the critical velocity as the point at which this local velocity reaches the sound velocity. 
Eventually, the critical velocity within the leading order correction can be obtained from the following equation:
\begin{widetext}
\begin{align}
\frac{M}{24 L^4 \Lambda^2} \left[ L^5 (12 + 7 M^2) \Lambda + 256 M^2 l^5 (-l + \Lambda) - 16 L^3 M^2 l^2 (8 l + \Lambda)+ 128 L M^2 l^4 (-6 l + 5 \Lambda)  \right.\nonumber\\
\left.+ 32 L^2 M^2 l^3 (-22 l + 13 \Lambda)+ 2 L^4 l \left( 12 \Lambda + M^2 \left( 32 l - 17 \Lambda \right) \right) \right]=1
\label{eq:hosei_eq}
\end{align}
\end{widetext}
with $\Lambda \equiv \sqrt{l(L+l)}$.
The result including the leading order correction is shown by the dotted curve in Fig.\ref{fig:result_1}. 
Here, the fitting parameter $l$, the distance from the edge of the plate, is set to $l=l_\xi$ as before. 
Compared to the result before applying the correction (red solid line), the critical velocity shifts lower values as a whole and agrees better with the numerical results over a wider range of $L$. 
In other words, in the region where the obstacle size is small, it is necessary to incorporate the effect of compressibility due to the increased background velocity.
We repeat the similar calculations by including the higher-order corrections; the results for the order expansion up to $M^{30}$ in Eq.\eqref{eq:pot_tenkai} are also shown in Fig.\ref{fig:result_1}. 
It is remarkable that the suppression of $U_c/c_s$ for $L<20$ is enhanced further.  
In the region where the plate size is on the order of the healing length, $L/\xi \sim 1$, there is a noticeable discrepancy between the numerical results and the critical velocity derived from the complex velocity potential.
There, the quantum pressure cannot be negligible since the scale of generated vortex core $\sim \xi$ has the same extent of the obstacle size. 

\subsection{Plate with finite thickness}
Finally, we briefly show the impact of the finite thickness of the plate-shaped obstacle on the critical velocity.
The obstacle potential $V_\textrm{b}(x,y)$ with a finite thickness $d$ is set as
\begin{equation}
    V_\textrm{b}(x,y)=\begin{cases}V_0 & |y| \leq d\  \textrm{and}\  |x|\leq L/2\\0 & \textrm{otherwise}\end{cases},
\end{equation}
and we follow the similar numerical procedure in Sec.\ref{sec:2a} to calculate $U_c$.
\begin{figure*}[ht]
    \centering
    \includegraphics[width=\linewidth]{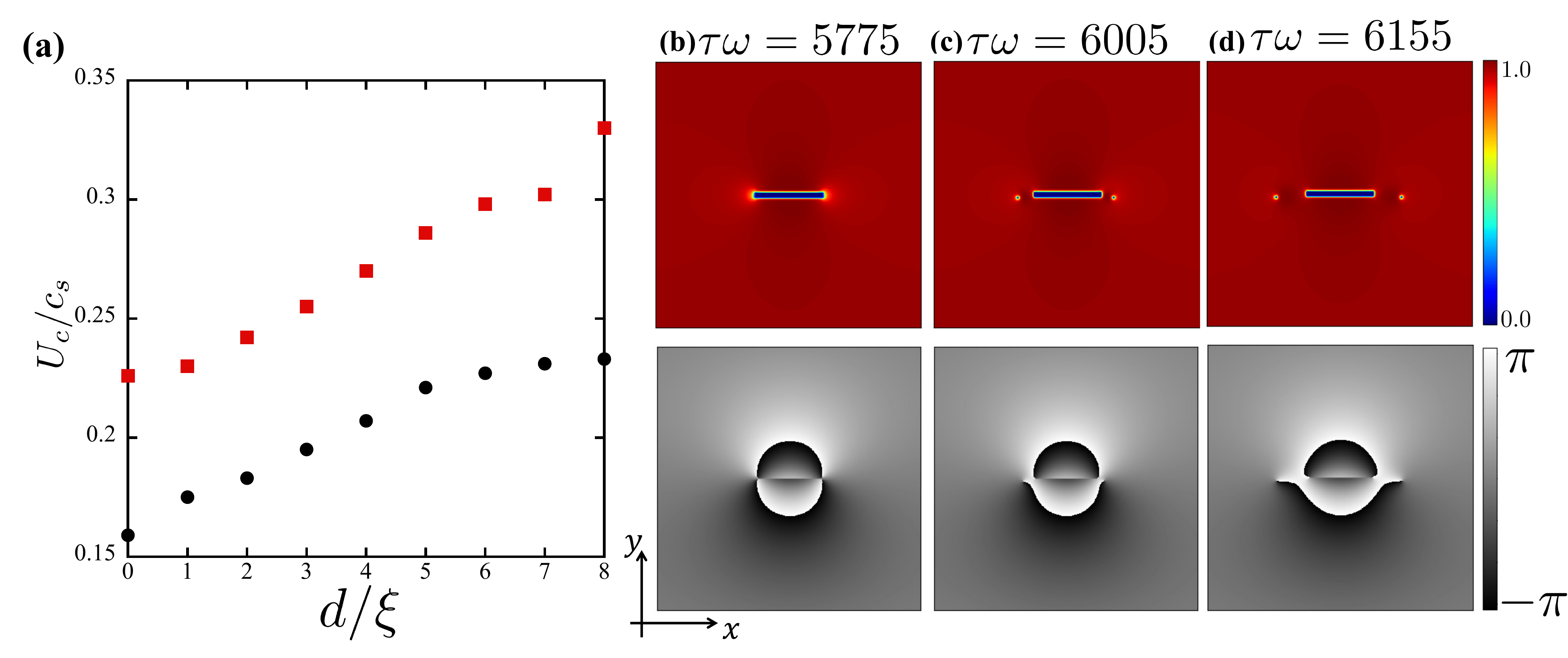}
    \caption{[Panel(a)] The dependence of the critical velocity on the thickness $d/\xi$ of a box-shaped obstacle, obtained by numerical calculations. The black circles represent the results for $L=40$, while the red squares represent the results for $L=80$. The critical velocity monotonically increases as the plate becomes thicker.
    [Panel(b)-(d)] Snapshots of the density profile (top) and the phase profile (bottom) of the condensate wave function by the imaginary time development for $L=80\xi$, $d=5\xi$ and $U=0.221c_s$.
    The spatial region of the plot is $-160\xi<x<160\xi$ and $-160\xi<y<160\xi$.
    The vortices are emitted from the edge of the plate in the direction of the plate, and these vortices disappear at the boundary.
    }
    \label{fig:box_result}
\end{figure*}
Figure \ref{fig:box_result}(a) shows the numerical results for the dependence of critical velocity on thickness when the plate length is $L=40,80$. 
To investigate clearly the trends due to the thickness, we examine its effect by focusing solely on regions with large plate sizes where compressibility is not significant.
From these results, it can be confirmed that the critical velocity increases as the thickness increases. 
This tendency can be qualitatively understood from the shape of the obstacle. 
First, in the case of an infinitely thin plate, the direction of the superflow passing near the edge of the plate must rotate $180\tcdegree$ from the front to the back of the plate to satisfy the equation of continuity. This flow distribution is advantageous for vortex formation. 
On the other hand, in the case of a box-shaped obstacle, the superfluid flows around the corner of the box by $90\tcdegree$. 
The formation of vortices from the corners of the box-shaped obstacle can also be observed in Figs.\ref{fig:box_result}(b)-(d).
As a result, the local velocity around the corner is slower than that around the plate edge under the same background velocity $U$, resulting a higher critical velocity for the former.

\section{Conclusion and discussion}\label{sec:conclusion}
In this study, we calculated the critical velocity for vortex shedding in the wake of a plate-shaped obstacle moving through a uniform superfluid.
Our method based on the potential flow theory succeeded in predicting the explicit size dependence of the critical velocity observed in the numerical simulation. 
When the size $L$ of the plate-shaped obstacle is large, the critical velocity exhibits a power law of $v_c/c_s\sim (L/\xi)^{-1/2}$, more precisely described by Eq.\eqref{eq:vc_p1} within the incompressible approximation.
Furthermore, it was demonstrated that the critical velocity increases for smaller obstacles, which motivates us to consider the compressibility. 
By adding a correction term for potential flow in terms of Mach number $M=U/c_s$, the critical velocity can be determined by solving Eq.\eqref{eq:hosei_eq}. 
The analytical critical velocity obtained in this way shows good agreement with the results from numerical calculations. 
In the region where the plate size is on the order of the healing length, $L/\xi \sim 1$, a discrepancy was observed between the numerical results and the critical velocity derived from the complex velocity potential. 
This suggests that the present analysis is not valid in such regions, where the quantum pressure term should be dealt with in more details over our phenomenological analysis.
Finally, the critical velocity for a plate-shaped potential with thickness, was determined using numerical calculations. 
As shown in Fig.\ref{fig:box_result}, the increase in critical velocity with the thickness of the box-shaped obstacle was confirmed.
This is attributed to the reduction in the flow-around angle compared to the plate-shaped obstacle.
In experiments, optical lasers will be used to replicate a plate-shaped obstacle, giving some thickness caused by the finite width of the laser beam. 
Our treatment would be correct if the actual thickness of the potential is sufficiently small compared to the healing length.

Our result is important to understand the quantum counterpart of the Reynolds similitude of classical hydrodynamics.
A characteristic of quantum nature in the superfluid wake was described through the superfluid Reynolds number \cite{Nore1997-eb,Nore2000-zt,Volovik2003-tj,Reeves2015-fi,Takeuchi2024-ki,Christenhusz2024-by}, defined by substituting the quantized circulation $\kappa$ for the kinematic viscosity $\nu$ in the conventional Reynolds number $UL/\nu$ with the characteristic velocity $U$ and size $L$. 
Later, the velocity correction $U-U_c$ was introduced, since the quantum vortices can be generated when the velocity $U$ exceeds a critical value $U_c$ \cite{Reeves2015-fi}.
It is worth investigating whether the classical universality is still possible in such microscopic regimes when using Reynolds numbers with accurately predicted critical velocities.
In future work, we use the analytically derived critical velocity to perform a systematic analysis for the dynamics of superfluid wake flows based on the superfluid Reynolds number.
\bibliographystyle{apsrev4}
\bibliography{vc_pl}
\end{document}